\documentclass[prc,floatfix,groupedaddress,nofootinbib,showpacs,preprintnumbers,
amsmath,amssymb,amsfonts,superscriptaddress,widetable] {revtex4}
\usepackage{graphicx}
\usepackage{dcolumn}
\usepackage{mathrsfs}
\usepackage{bm}

\usepackage{graphicx}
\usepackage{dcolumn}
\usepackage{bm}
\usepackage[usenames]{color}
\begin{document}

\title{Neutron skins and neutron stars}
\author{F. J. Fattoyev}
\email{Farrooh_Fattoyev@tamu-commerce.edu} \affiliation{Department
of Physics, Florida State University, Tallahassee, Florida 32306,
USA} \affiliation{Department of Physics and Astronomy, Texas A\&M
University-Commerce, Commerce, Texas 75429-3011, USA}
\author{J. Piekarewicz}
\email{jpiekarewicz@fsu.edu} \affiliation{Department of Physics,
Florida State University, Tallahassee, Florida 32306, USA}

\date{\today}
\begin{abstract}
 {\bf Background:} The neutron skin of a heavy nucleus as
 well as many neutron-star properties are highly sensitive
 to the poorly constrained density dependence of the
 symmetry energy.
 {\bf Purpose:} To provide for the first time meaningful
 theoretical errors and to assess the degree of correlation
 between the neutron-skin thickness of ${}^{208}$Pb and
 several neutron-star properties.
 {\bf Methods:} A proper covariance analysis based on
 the predictions of an accurately-calibrated relativistic
 functional {\sl ``FSUGold''}  is used to quantify theoretical
 errors and correlation coefficients.
 {\bf Results:} We find correlation coefficients of nearly
 one (or minus one) between the neutron-skin thickness
 of ${}^{208}$Pb and a host of observables of relevance to
 the structure, dynamics, and composition of neutron stars.
 {\bf Conclusions:}  We suggest that a follow-up PREX
 measurement, ideally with a $0.5\%$ accuracy, could
 significantly constrain the equation of state of
 neutron-star matter.
\end{abstract}
\maketitle

\section{Introduction}
\label{intro}

The equation of state (EOS) of neutron-rich matter plays a crucial
role in understanding interesting phenomena in both nuclear physics
and astrophysics, such as the limits of nuclear existence, the
dynamics of heavy ion collisions, the structure of neutron stars,
and the mechanism of core-collapse supernovae. A particularly
important property of the EOS is the nuclear symmetry energy, whose
value is well constrained by the ground-state properties of finite
nuclei only near saturation density~\cite{Furnstahl:2001un}.  Given
that the symmetry energy reflects the increase in the energy of the
system as protons are turned into neutrons (or viceversa), many
nuclear and astrophysical observables are sensitive to its density
dependence. For example, the EOS of neutron-rich matter is the sole
feature responsible for the structure of neutron stars. Moreover,
the neutron-skin thickness of heavy nuclei---a system that is 18
orders of magnitude smaller than a neutron star---is also sensitive
to the symmetry energy. This is because the slope of the symmetry
energy is related to the pressure exerted by the neutrons in
creating both a neutron-rich skin and the neutron-star
radius~\cite{Horowitz:2001ya,Horowitz:2000xj}.  Although
considerable effort has been devoted to the understanding of the EOS
of isospin asymmetric matter and its possible impact on both
laboratory and observational data~\cite{Steiner:2004fi,Li:2008gp},
our knowledge of the density dependence of the symmetry energy
remains incomplete. Whereas some of the laboratory/observational
data is relatively easy to collect, most requires great perseverance
and ingenuity.  Thus, in an effort to identify observables sensitive
to the density dependence of the symmetry energy and to establish
correlations among them, we rely on a powerful and systematic
covariance analysis. Such a systematic statistical analysis may be
used to attach meaningful uncertainty estimates to theoretical
predictions~\cite{PhysRevA.83.040001}.  Moreover, it can be used to
identify observables that, although at present may be beyond
experimental/observational reach, display a strong sensitivity to
the symmetry energy.  Finally, through such a covariance analysis
one can quantify the degree of correlation between various
observables. It is the aim of the present contribution to establish
for the first time quantitative correlations between the
neutron-skin thickness of $^{208}$Pb and a variety of neutron-star
observables.

In a recent publication we developed a covariance analysis within a
class of relativistic mean-field models~\cite{Fattoyev:2011ns}.
Starting from a $\chi^2$-minimization procedure, we presented a
step-by-step implementation of the statistical approach that provided
quantitative uncertainties in our theoretical predictions as well as
robust correlations among physical observables. In the present
contribution we extend such a study to: (a) quantify the degree of
correlation between the neutron-skin thickness of $^{208}$Pb and
several neutron-star properties and (b) provide meaningful theoretical
uncertainties that arise from our incomplete knowledge of the density
dependence of the symmetry energy.  We have selected the neutron-skin
thickness of $^{208}$Pb as it represent a laboratory observable of
paramount importance in constraining the density dependence of the
symmetry energy. Indeed, the strong correlation between the
neutron-skin thickness of $^{208}$Pb and the slope of the symmetry
energy at saturation density is a well established
fact~\cite{Brown:2000,Furnstahl:2001un,
Centelles:2008vu,Centelles:2010qh,Vidana:2009is}. Further, the highly
anticipated Lead Radius Experiment (PREX) has just provided the first
model-independent evidence of the existence of a significant neutron
skin in ${}^{208}$Pb~\cite{Abrahamyan:2012gp}.  Building on the
strength of the enormously successful parity-violating program at the
Jefferson Laboratory, PREX used parity-violating electron scattering
to determine the neutron-skin thickness of $^{208}$Pb to be:
\begin{equation}
 R_{\rm skin}\!\equiv\!R_{n}\!-\!R_{p}\!=\!0.33^{+0.16}_{-0.18}~{\rm fm} \;,
 \label{PREX}
\end{equation}
where $R_{n}(R_{p})$ denotes the root-mean-square neutron(proton)
radius. Although PREX achieved the systematic control required to
perform this challenging experiment, unforeseen technical problems
resulted in time losses that significantly compromised the statistical
accuracy of the measurement. Thus, rather than achieving the original
goal of a 1\% error in the neutron radius, PREX had to settle for an
error almost three times as large. Even so, the PREX measurement has
already been used to impose some (mild) constraints on
accurately-calibrated models of
nuclear-structure~\cite{Piekarewicz:2012pp}. Moreover, the---already
approved---follow-up PREX measurement designed to achieve the original
1\% goal will be of great value, especially if the unexpected large
central value remains unchanged.

Since the strong correlation between the neutron-skin thickness of
$^{208}$Pb and the pressure of pure neutron matter near saturation
density was first inferred~\cite{Brown:2000,Furnstahl:2001un}, a large
number of additional correlations between $R_{\rm skin}$ and several
neutron-star properties have been uncovered~\cite{Steiner:2004fi}.
These include: (a) the crust-to-core transition
density~\cite{Horowitz:2000xj}, (b) neutron-star
radii~\cite{Horowitz:2001ya,Carriere:2002bx}, (c) threshold density
at the onset of the direct Urca process~\cite{Horowitz:2002mb}, and
(d) the crustal moment of inertia~\cite{Steiner:2004fi,Fattoyev:2010tb},
among others.  Although most of these correlations have been
previously established through a systematic variation of a previously
calibrated model, in the present manuscript we attempt for the first
time to reliably quantify all these correlations via a legitimate
covariance analysis. A covariance analysis represents the least
biased and most exhaustive tool to uncover correlations between
physical observables~\cite{Reinhard:2010wz,Fattoyev:2011ns}.

The manuscript has been organized as follows. In Sec.~\ref{Formalism}
we briefly review the formalism required to implement the covariance
analysis developed recently in Ref.~\cite{Fattoyev:2011ns}. In
Sec.~\ref{Results} we present physical arguments in favor of the
expected correlations between the neutron-skin thickness of $^{208}$Pb
and several neutron-star properties. Properly estimated theoretical
errors and correlation coefficients are also presented in this
section. Finally, conclusions and suggestions for future work are
summarized in Sec.~\ref{Conclusions}.

\section{Formalism}
\label{Formalism}

The covariance analysis mentioned above will be implemented using the
predictions from the accurately calibrated FSUGold
model\,\cite{Todd-Rutel:2005fa}.  The interacting Lagrangian density
for the model---with its predictions generated at the relativistic
mean-field (RMF) level---is given by
\begin{eqnarray}
{\mathscr L}_{\rm int} &=& \bar\psi \left[g_{\rm s}\phi   \!-\!
         \left(g_{\rm v}V_\mu  \!+\!
    \frac{g_{\rho}}{2}{\mbox{\boldmath $\tau$}}\cdot{\bf b}_{\mu}
                               \!+\!
    \frac{e}{2}(1\!+\!\tau_{3})A_{\mu}\right)\gamma^{\mu}
         \right]\psi \nonumber \\
                   &-&
    \frac{\kappa}{3!} (g_{\rm s}\phi)^3 \!-\!
    \frac{\lambda}{4!}(g_{\rm s}\phi)^4 \!+\!
    \frac{\zeta}{4!}   g_{\rm v}^4(V_{\mu}V^\mu)^2 +
   \Lambda_{\rm v} g_{\rho}^{2}\,{\bf b}_{\mu}\cdot{\bf b}^{\mu}
           g_{\rm v}^{2}V_{\nu}V^\nu\;.
 \label{LDensity}
\end{eqnarray}
This model contains an isodoublet nucleon field ($\psi$) interacting
via the exchange of two isoscalar mesons, the scalar sigma ($\phi$)
and the vector omega ($V^{\mu}$), one isovector meson, the rho (${\bf
b}^{\mu}$), and the photon ($A^{\mu}$). In order to improve the
standing of the model the Lagrangian density is supplemented by scalar
and vector self-interactions. In particular, the scalar
self-interaction (with coupling constants $\kappa$ and $\lambda$) is
responsible for reducing the compression modulus of nuclear matter
from the unrealistically large value of $K\!=\!545$~MeV obtained with
the original Walecka model~\cite{Walecka:1974qa,Serot:1984ey} to about
$K\!=\!230$~MeV. Such a significantly lower value for the compression
modulus is demanded by measurements of giant monopole resonances in
medium to heavy
nuclei~\cite{Youngblood:1999,Piekarewicz:2001nm,Piekarewicz:2002jd,
Colo:2004mj}.  The quartic isoscalar-vector self-interaction is
responsible for softening the EOS at high densities. Indeed, by tuning
$\zeta$ one can generate different limiting neutron-star masses
without modifying the behavior of the EOS around saturation
density~\cite{Mueller:1996pm}.  As such, $\zeta$ is fairly insensitive
to laboratory observables and must be constrained from astrophysical
observations.  Finally, the mixed quartic vector interaction (as
described by parameter $\Lambda_{\rm v}$) was introduced to modify the
density dependence of symmetry energy~\cite{Horowitz:2001ya}, which is
fairly stiff in the original Walecka model. By calibrating the
parameters of the model using both ground-state properties of finite
nuclei as well as their collective
excitations~\cite{Todd-Rutel:2005fa}, the FSUGold model has been
fairly successful when compared against theoretical, experimental, and
observational constraints~\cite{Piekarewicz:2007dx}. However, as more
observational and experimental data become available---such as the
recently reported 2-solar mass neutron
star~\cite{Demorest:2010bx}---further refinements may be
required~\cite{Fattoyev:2010mx} (for example by tuning $\zeta$).
Yet, for the purpose of this contribution we will be satisfied with
the use of FSUGold model as the basis for the covariance analysis.

The covariance analysis follows closely the approach outlined in
Refs.~\cite{Reinhard:2010wz,Fattoyev:2011ns} which, in turn, is
based on the comprehensive text by Brandt~\cite{Brandt:1999}.
The derivation, implementation, and power of the formalism was
illustrated in our recent study using two relativistic mean-field
models~\cite{Fattoyev:2011ns}. Thus, in the present contribution
we only offer a brief summary.
The covariance analysis starts with the definition of the
quality measure $\chi^2$. That is,
\begin{equation}
 \chi^{2}({\bf p}) \equiv \sum_{n=1}^{N}
 \left(\frac{\mathcal{O}_{n}^{\rm (th)}({\bf p})-
 \mathcal{O}_{n}^{\rm (exp)}}
 {\Delta\mathcal{O}_{n}}\right)^{2} \;,
 \label{ChiSquare}
\end{equation}
where $N$ denotes the total number of selected observables that will
be employed in the calibration procedure. Each of the observables,
$\mathcal{O}_{n}^{\rm (exp)}$, is assumed to have been determined
experimentally with an accuracy of $\Delta\mathcal{O}_{n}$, which acts
as a weight factor in the quality measure. In addition, each of the
$N$ observable is computed within the model $\mathcal{O}_{n}^{\rm
(th)}({\bf p})$ as a function of the $F$ model-parameters ${\bf
p}\!=\!(p_{1},\ldots,p_{F})$. The calibration procedure terminates
when a set of optimal parameters ${\bf p}_{0}$ are found that minimize
the quality measure.

Given that our main goal is not the minimization procedure, but rather
the estimation of theoretical uncertainties and the assessment of
correlations among observables, we use a set of nuclear properties
generated directly from the FSUGold model.  This guarantees that the
quality measure so defined is minimized. In particular, the following
set of $N\!=\!8$ observables have been selected as input for $\chi^2$:
(1) the saturation density $\rho_{0}$, (2) the binding energy per
particle of symmetric nuclear matter at saturation density
$\varepsilon_{0}$, (3) the binding energy per particle of symmetric
nuclear matter at twice saturation density $\varepsilon(2\rho_{0})$,
(4) the compression modulus of symmetric nuclear matter $K_{0}$, (5)
the effective (Dirac) mass of symmetric nuclear matter at saturation
density $M_{0}^{\star}$, (6) the symmetry energy $\tilde{J}$ evaluated
at a sub-saturation density of $\rho\!\approx\!0.1\,{\rm fm}^{-3}$,
(7) the slope of the symmetry energy at saturation density $L$, and
(8) the maximum neutron-star mass $M_{\rm max}$. We attach a 2\%
uncertainty to all of the observables---except for the case of the
slope of the symmetry energy $L$ where the significantly larger value
of 20\% is assumed. Such a large uncertainty reflects our poor
understanding of the density dependence of the symmetry energy.
Note that our choice of $\tilde{J}$, namely, the symmetry energy at
sub-saturation density rather than at saturation density, follows
from the fact that heavy nuclei constrain the symmetry energy at a
density which is intermediate between the center and the surface of
the nucleus.  Indeed, at such a sub-saturation density the theoretical
uncertainties appear to be minimized~\cite{Furnstahl:2001un}.

Having obtained the optimal parameter set ${\bf p}_{0}$ through the
minimization of the quality measure, one can the proceed to compute
and diagonalize the symmetric matrix of second derivatives. This
matrix contains all the information about the behavior of the $\chi^2$
function around the minimum. That is,
\begin{equation}
 \chi^2({\bf p}) -\chi^{2}({\bf p}_{0})
 \equiv \Delta\chi^2({\bf x}) =
 {\bf x}^{T}{\hat{\mathcal M}}\,{\bf x} = {\bm\xi}^{T}
   {\hat{\mathcal D}}{\bm\xi}=\sum_{i=1}^{F}
   \lambda_{i}\xi_{i}^{2} \;,
 \label{Taylor2}
\end{equation}
where
\begin{equation}
  x_{i} \equiv \frac{({\bf p}-{\bf p}_{0})_{i}}{({\bf p}_{0})_{i}} \;
 \label{xDef}
\end{equation}
are scaled dimensionless variables, ${\hat{\mathcal M}} =
{\hat{\mathcal A}}{\hat{\mathcal D}} {\hat{\mathcal A}^{T}}$, and
${\bm\xi}\!=\!{\hat{\mathcal A}^{T}}{\bf x}$ are dimensionless
variables in a rotated basis. Here ${\hat{\mathcal A}}$ is the
orthogonal matrix whose columns are composed of the normalized
eigenvectors and ${\hat{\mathcal D}}\!=\!{\rm diag}
(\lambda_{1},\ldots,\lambda_{F})$ is the {\sl diagonal} matrix of
eigenvalues.  Although the full covariance analysis may be carried out
without the need to diagonalize the matrix of second derivatives
${\hat{\mathcal M}}$, doing so is both simple and illuminating.  For
example, the small oscillations around the $\chi^2$-minimum may be
represented as a collection of $F$ uncoupled harmonic
oscillators. Doing so readily identifies the {\sl ``stiff''} and {\sl
``soft''} modes in parameter space.  In particular, the soft modes
indicate the linear combination of parameters that are poorly
constrained by the choice of observables included in the quality
measure. In turn, this suggets the kind of additional physical
observables that are required to further constrain the model. In the
particular case of FSUGold, the softest direction is dominated by the
{\sl ``in-phase motion''} of the isovector coupling constants
$g_{\rho}$ and $\Lambda_{\rm v}$~\cite{Fattoyev:2011ns}. This
particular linear combination remains largely unconstrained because of
our poor knowledge of the density dependence of the symmetry energy.

To obtain meaningful theoretical uncertainties as well as to assess
the degree of correlation between two observables, one must compute
the statistical covariance of two observables $A$ and $B$. Assuming
that the model parameters are distributed according to quality measure
as $\exp\left(-{\bf x}^{T}{\hat{\mathcal M}}\,{\bf x}/2\right)$, the
covariance of $A$ and $B$ may be written as follows:
\begin{equation}
 {\rm cov}(A,B) = \sum_{i,j=1}^{F}
 \frac{\partial A}{\partial x_{i}}
  (\hat{{\mathcal M}}^{-1})_{ij}
 \frac{\partial B}{\partial x_{j}} =
 \sum_{i=1}^{F}
 \frac{\partial A}{\partial \xi_{i}}
 \lambda_{i}^{-1}
 \frac{\partial B}{\partial \xi_{i}} \;.
 \label{Covariance}
\end{equation}
Note that the variance $\sigma^{2}(A)$ of a given observable $A$
is simply given by $\sigma^2(A)\!=\!{\rm cov}(A,A)$. Also note
that, all other things being equal, the $ {\rm cov}(A,B)$ is dominated
by the softest direction. Finally, the Pearson product-moment
correlation coefficient---or correlation coefficient for
simplicity---is defined as:
\begin{equation}
 \rho(A, B) = \frac{{\rm cov}(A,B)}
 {\sqrt{ {\rm var}(A) {\rm var}(B) }} \;.
 \label{Correlation}
\end{equation}
Further details on the covariance analysis may be found in
Refs.~\cite{Reinhard:2010wz,Fattoyev:2011ns,Brandt:1999}.

\section{Results}
\label{Results}

The main goal of this contribution is to use a covariance analysis to
reliably assess the correlation between the neutron-skin thickness of
${}^{208}$Pb ($R_{\rm skin}$) and a large number of observables of
relevance to the structure, dynamics, and composition of neutron
stars.  Moreover, we rely on a covariance analysis to attach
meaningful uncertainty estimates to our theoretical predictions. In
particular, the FSUGold model predicts the following neutron-skin
thickness of ${}^{208}$Pb with properly computed theoretical
errors~\cite{Fattoyev:2011ns}:
\begin{equation}
 R_{\rm skin} = (0.2069 \pm 0.0366)\,{\rm fm}\,\,[17.698\%] \;.
 \label{RskinFSU}
\end{equation}
Note that this prediction fits comfortably within the recently
determined value by the PREX collaboration [see Eq.~(\ref{PREX})].
However, there is a significant difference in the central value that
may prove very interesting if the follow-up PREX measurement can
significantly reduce the error bars without dramatically affecting the
central value.

Given the large number of observables that will be discussed, the
present section has been divided into several subsections. In
particular, each subsection motivates the connection of a given
neutron-star observable to $R_{\rm skin}$ and discusses its relevance
in constraining the density dependence of the symmetry energy in
regions that may be inaccessible to laboratory experiments.

\subsection{Pure Neutron Matter}
\label{Results:PNM}

\begin{figure}[ht]
\vspace{-0.05in}
\includegraphics[width=0.6\columnwidth,angle=0]{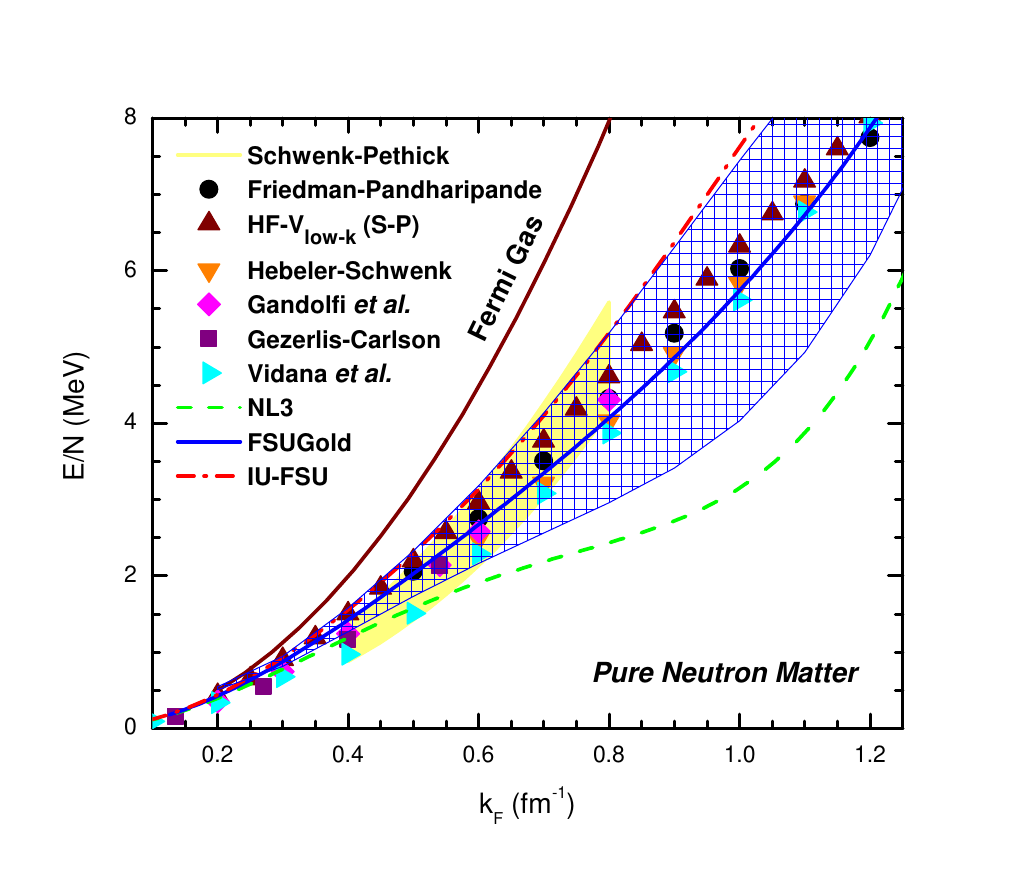}
\caption{(Color online)  Theoretical uncertainties in the energy per
neutron as a function of the Fermi momentum for pure neutron matter.
The shaded blue area are the $1\sigma$ theoretical uncertainty
associated to the FSUGold model.} \label{Fig1}
\end{figure}

Assuming the validity of General Relativity, the equation of state of
cold, catalyzed neutron-rich matter is the sole ingredient that
determines the structure of spherical neutron stars in hydrostatic
equilibrium. As such, one of the most stringent constraints on the
structure of neutron stars comes from the EOS of pure neutron matter
(PNM). Although PNM remains a theoretical construct, enormous progress
has been made in constraining its equation of state at low
densities. Interestingly enough, most of the progress in this area has
been driven by remarkable advances in cold-atom experiments that make
possible to study the universal behavior of resonant Fermi gases at
the unitary limit of infinite scattering length.  Even though the
appreciable effective range of the neutron-neutron interaction
invalidates some of the powerful arguments associated with resonant
Fermi gases, model independent results have been obtained for dilute
neutron matter~\cite{Schwenk:2005ka}.  Moreover, the full power of
quantum Monte Carlo methods has been used to extend the calculations
to higher densities~\cite{Gezerlis:2009iw,Gezerlis:2011ai}.  By doing
so, the calculated EOS---which matches smoothly to the analytic
results---provides important constraints on nuclear density
functionals. Indeed, the impact of such constraints can be clearly
assessed in Fig.~\ref{Fig1} where the equation of state of pure
neutron matter for a variety of microscopic approaches (see
Refs.~\cite{Gezerlis:2009iw,Gezerlis:2011ai} and references therein)
is displayed alongside the predictions from three relativistic
functionals. Note that although not included in the calibration
procedure, the EOS predicted by the FSUGold interaction (solid blue
line) appears consistent with most of the microscopic approaches. Also
note that the required {\sl ``softening''} of the EOS relative to the
more traditional RMF models (such as NL3) was generated by
constraining the calibration procedure by the dynamics of nuclear
collective modes~\cite{Todd-Rutel:2005fa}.

\begin{figure}[ht]
\vspace{-0.05in}
\includegraphics[width=0.35\columnwidth,angle=0]{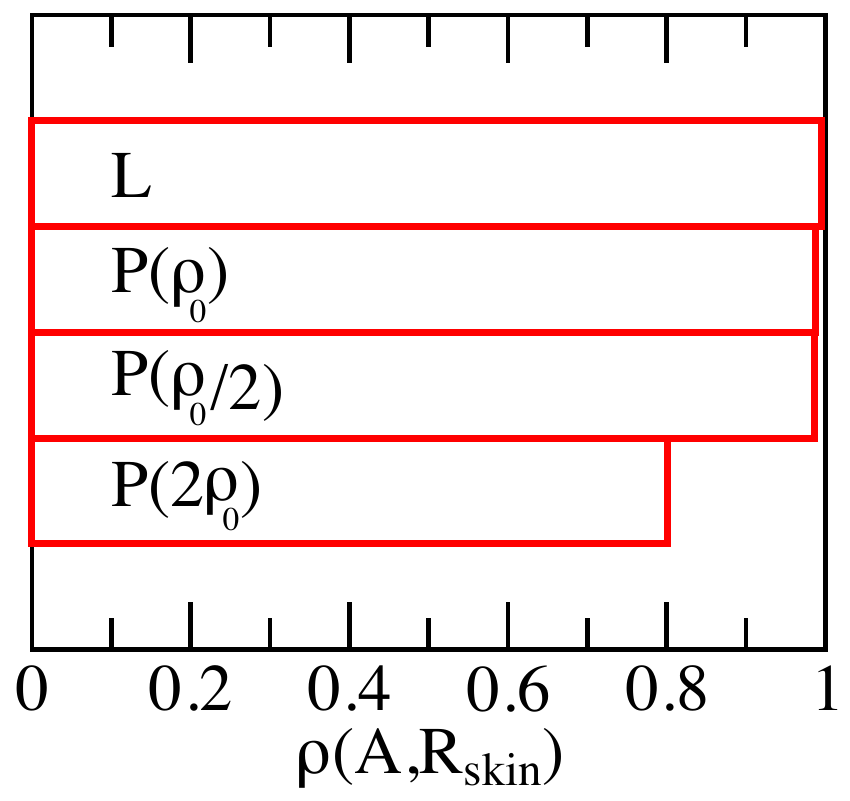}
\caption{(Color online) The correlation coefficient between the
neutron-skin thickness of $^{208}$Pb and the slope of the
symmetry energy $L$ and the pressure of pure neutron matter
at three values of the density.}
\label{Fig2}
\end{figure}

The wide (blue) band in Fig.~\ref{Fig1} represents the $1\sigma$
uncertainty in the predictions of the FSUGold model.  That the
$1\sigma$-band is so wide reflects our poor understanding of the
density dependence of the symmetry energy (recall that a 20\%
uncertainty was assumed for the slope of the symmetry energy $L$).  In
order to reduce such a large uncertainty one must constrain $L$---or
equivalently the pressure of PNM at saturation---through an accurate
measurement of the neutron radius in ${}^{208}$Pb.  Although the
correlation between the neutron radius of ${}^{208}$Pb and the slope
of the symmetry energy $L$ is by now very well established, only
recently has a statistically meaningful correlation coefficient
between such quantities been
computed~\cite{Reinhard:2010wz,Fattoyev:2011ns}.  We have now extended
such a covariance analysis to explore the correlation between the
neutron-skin thickness of ${}^{208}$Pb and the pressure of PNM at
half, one, and two times nuclear-matter saturation density (see
Table~\ref{Table2}  and Fig.~\ref{Fig2}).

  \begin{table}[h]
  \begin{tabular}{|c|r|c|}
    \hline
     $A$  & $\langle A\rangle\pm\Delta A\hfil$ & $\rho(A,R_{\rm skin})$ \\
    \hline
     $L({\rm MeV})$   & $(60.5152 \pm 12.1011)\, [19.997\%]$ & $0.9952$ \\
     $P(\rho_{0})$       & $(3.1842 \pm 0.6349)\,[19.940\%]$ & $0.9882$ \\
     $P(\rho_{0}/2)$   & $(0.4874 \pm 0.1721)\,[35.304\%]$ & $0.9861$ \\
     $P(2\rho_{0})$     & $(21.8569 \pm 1.2735)\,[5.827\%]$ & $0.8016$ \\
    \hline
  \end{tabular}
 \caption{Theoretical errors associated to the predictions of the
                FSUGold model for the slope of the symmetry energy
                and the pressure of pure neutron matter
                (in units of ${\rm MeV}\,{\rm fm}^{-3}$) at three
                different densities.}
 \label{Table2}
 \end{table}

Given that the symmetry energy is to a very good approximation equal
to the difference in energy between PNM and symmetric nuclear matter,
the slope of the symmetry energy $L$ is closely related to the
pressure of PNM at saturation density, {\sl i.e.,}
$P(\rho_{0})\!\approx\!\rho_{0}L/3$. Thus, perhaps not surprisingly,
we find correlation coefficients of almost one between $R_{\rm skin}$ and
both $L$ and $P(\rho_{0})$. Note, however, that the correlation
remains as strong for the pressure of PNM at half saturation density
$P(\rho_{0}/2)$---but deteriorates significantly for the pressure at
twice saturation density $P(2\rho_0)$. This last fact reflects the
insensitivity of finite-nuclei observables to the high-density
component of the equation of state.

\subsection{Neutron-Star Radii}
\label{Results:NSR}

The structure of neutron stars---particularly the mass-radius
relation---depends critically on the equation of state of neutron-rich
matter.  In particular, neutron-star radii provide stringent
constraints on the density dependence of the symmetry energy. Ideally,
one would measure neutron-star radii over a broad range of
masses. Indeed, measuring neutron-star radii $R(M)$ for a large range
of neutron star masses $M$ would allow one to directly deduce the
EOS~\cite{Lindblom:1992}. Unfortunately, whereas various neutron-star
masses are accurately known~\cite{Thorsett:1998uc}, a precise
determination of their radii does not yet exist.  Moreover,
constraining the EOS at low densities from the radii of low-mass
neutron stars may be difficult as these may be very rare. However, the
low-density EOS may be constrained from the neutron radii of heavy
nuclei as these contain similar
information~\cite{Horowitz:2001ya,Carriere:2002bx}.  This is because
the same pressure that is responsible for supporting a (low-mass)
neutron-star against gravitational collapse is also responsible for
the development of a neutron-rich skin in heavy nuclei. Indeed,
theoretical predictions suggest a strong correlation between these two
observables: {\sl the larger the neutron skin of a heavy nucleus, the
larger the stellar radius.}

\begin{figure}[ht]
\vspace{-0.05in}
\includegraphics[width=0.65\columnwidth,angle=0]{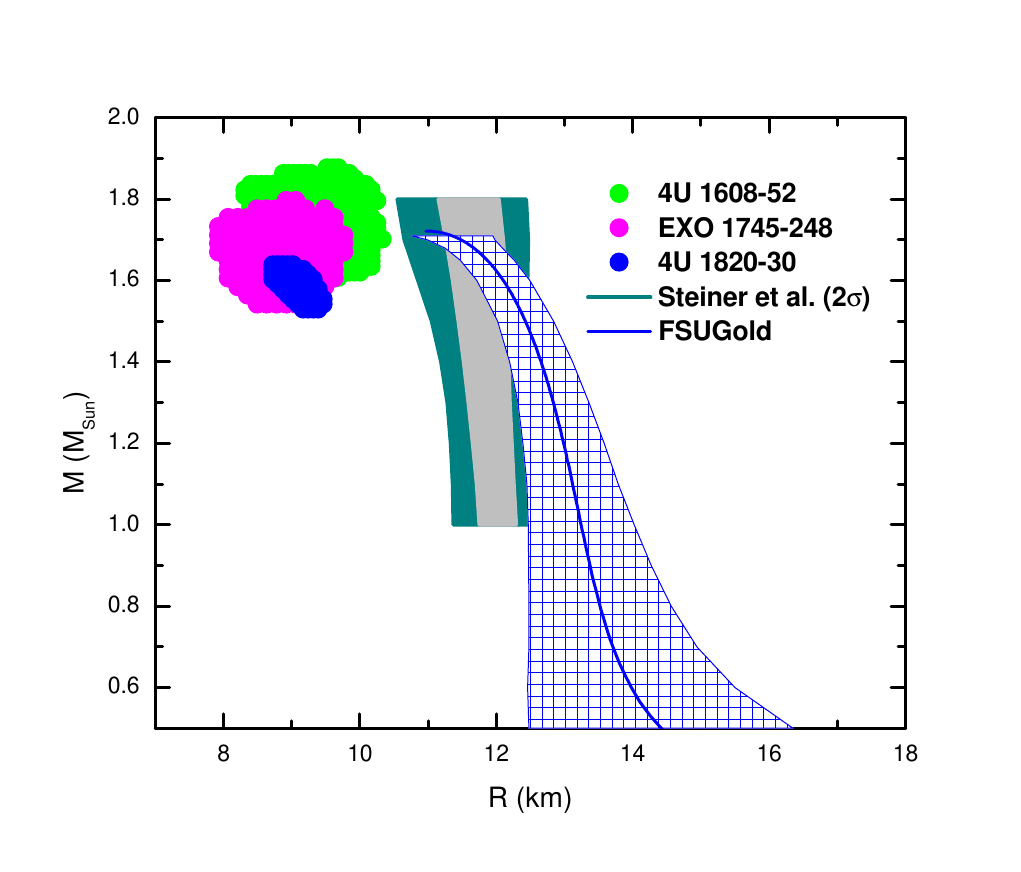}
\caption{(Color online) Theoretical predictions for the
mass-versus-radius relation of neutron stars calculated within
$1\sigma$ uncertainty in the FSUGold model. The  observational data
that suggest very small stellar radii represent 1$\sigma$ confidence
contours for the three neutron stars reported in
Ref.~\cite{Ozel:2010fw}. The two shaded areas that suggest larger
radii are 1$\sigma$ and 2$\sigma$ contours extracted from the
analysis of Ref.~\cite{Steiner:2010fz}.}
\label{Fig3}
\end{figure}

\begin{figure}[h]
\vspace{-0.05in}
\includegraphics[width=0.35\columnwidth,angle=0]{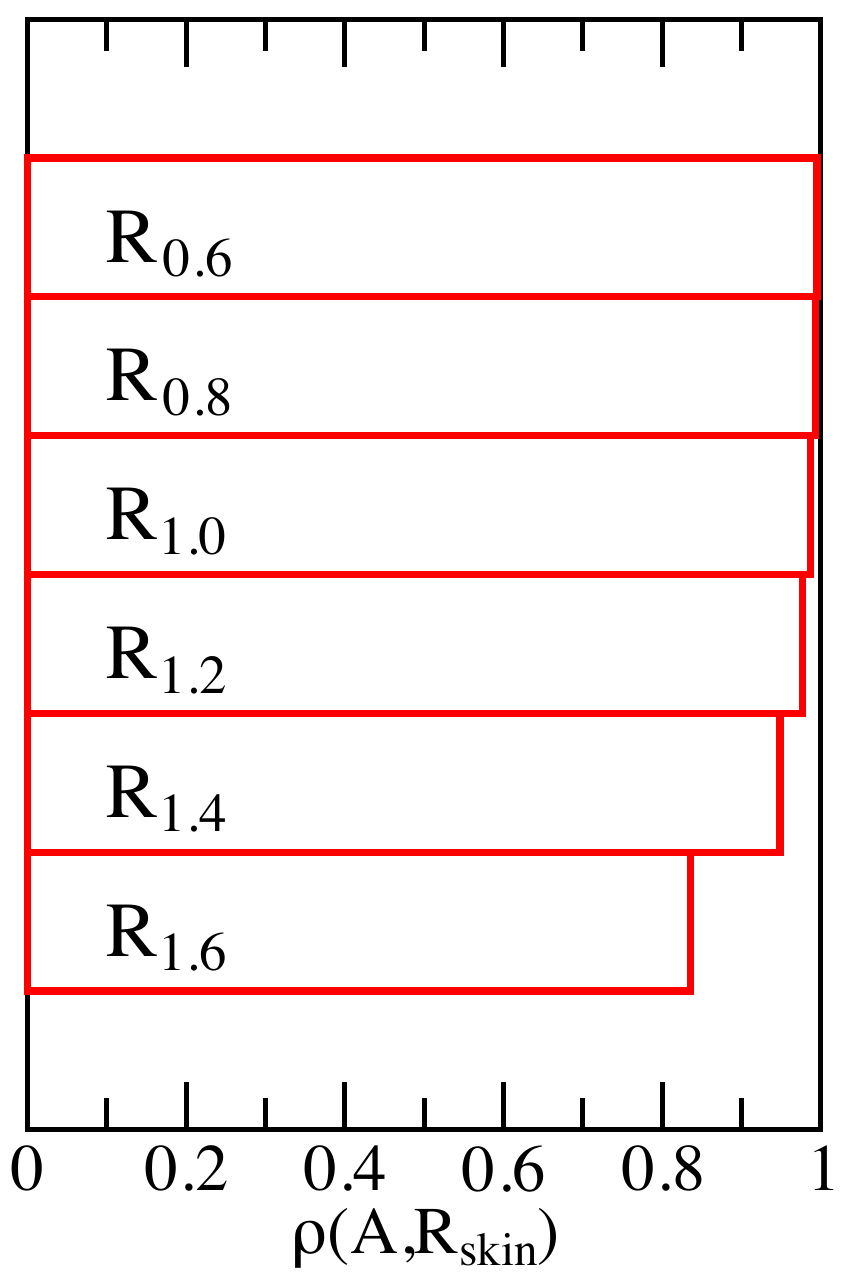}
\caption{(Color online) The correlation coefficient between the
neutron-skin thickness of $^{208}$Pb and the radius of neutron
stars of a variety masses. Note that the subscript indicates the
neutron-star mass in solar masses.}
\label{Fig4}
\end{figure}

To quantify the theoretical uncertainties in neutron-star radii and to
assess their correlation to $R_{\rm skin}$, we display in
Fig.\,\ref{Fig3} the {\sl Mass-vs-Radius} relation predicted by the
FSUGold model.  Although the determination of neutron-star radii from
observations of the luminosity and temperature is both challenging
and hindered by uncertainties in models of the stellar atmosphere (see
Ref.~\cite{Suleimanov:2009dp} and references therein) significant
advances in X-ray astronomy have allowed the simultaneous
determination of masses and radii from a systematic study of several
X-ray bursters~\cite{Ozel:2010fw, Steiner:2010fz}.  Results from such
studies are displayed in Fig.~\ref{Fig3} alongside the FSUGold
predictions.  We should note that while we believe that studies of
X-ray bursters will eventually become instrumental in constraining the
dense matter equation of state, at present they suffer from systematic
uncertainties~\cite{Steiner:2010fz}.  For example, the analysis from
Ref.~\cite{Ozel:2010fw} (occupying the upper left-hand corner of the
figure) suggests very small radii that are difficult to reconcile with
the predictions from relativistic mean-field
models~\cite{Fattoyev:2010rx}. In contrast, the more recent analysis
by Steiner, Lattimer, and Brown~\cite{Steiner:2010fz} suggests
significantly larger neutron-star radii that appear consistent with
the predictions of the model. Note that the wide (blue) band in
Fig.~\ref{Fig3} denotes our theoretical uncertainties in the
predictions of the stellar radii. We observe that whereas the stellar
radius is fairly well constrained for large-mass neutron stars, radii
of low-mass stars display large theoretical errors.  For example, the
radius of a $M\!=\!1.4M_{\odot}$ neutron star is determined with a
3.6\% uncertainty (Table\,\ref{Table4}).  Yet the error grows by a
factor of three for a $M\!=\!0.6M_{\odot}$ neutron star. Such behavior is
dominated by the large uncertainty in $L$. Whereas the radius of a
heavy neutron star is sensitive to the equation of state at both
intermediate and high densities, the radii of low-mass neutron stars
is sensitive to the range of densities probed in heavy nuclei. Indeed,
FSUGold predicts a central density in a $M\!=\!0.5M_{\odot}$ neutron
star of $0.21~{\rm fm}^{-3}$, only slightly higher than saturation
density.  Due to the large sensitivity of neutron stars radii to $L$,
a correlation between the neutron-skin thickness of ${}^{208}$Pb and
stellar radii has been established~\cite{Horowitz:2000xj,
Horowitz:2001ya, Carriere:2002bx}.  Moreover, we expect a stronger
correlation between $R_{\rm skin}$ and the radii of low-mass neutron
stars than between $R_{\rm skin}$ and the radii of massive stars. This
fact is clearly borne out by the results displayed in
Table\,\ref{Table4} and Fig.\,\ref{Fig4}.

  \begin{table}[h]
  \begin{tabular}{|c|r|c|}
    \hline
     $A$  & $\langle A\rangle\pm\Delta A\hfil$ & $\rho(A,R_{\rm skin})$ \\
    \hline
     $R_{0.6}$  & $(13.9785 \pm 1.5183)\,[10.862\%]$ & $0.9953$ \\
     $R_{0.8}$  & $(13.5204 \pm 1.0446)\, [7.726\%]$ & $0.9931$ \\
     $R_{1.0}$  & $(13.2439 \pm 0.7776)\, [5.872\%]$ & $0.9866$ \\
     $R_{1.2}$  & $(12.9864 \pm 0.5964)\, [4.593\%]$ & $0.9770$ \\
     $R_{1.4}$  & $(12.6568 \pm 0.4603)\, [3.637\%]$ & $0.9486$ \\
     $R_{1.6}$  & $(12.1038 \pm 0.3881)\, [3.206\%]$ & $0.8361$ \\
    \hline
  \end{tabular}
 \caption{Theoretical errors associated to the predictions of the
                FSUGold model for the radii (in km) of neutron stars
                of various masses. Note that the subscript indicates
                the neutron-star mass in solar masses.}
 \label{Table4}
 \end{table}

\subsection{Direct Urca Process}
\label{Results:DUP}

Neutron stars are born very hot (with a temperature of about
$10^{11}~{\rm K}$) and cool rapidly via neutrino emission
through the direct {\sl ``Urca'' process} involving neutron
beta decay and electron capture~\cite{Lattimer:1991ib,
Pethick:1991mk,Lattimer:2004pg,Page:2004fy}:
\begin{subequations}
 \begin{eqnarray}
  && n \rightarrow p + e^{-} + \bar\nu_{e}\;, \label{urca1}\\
  && e^{-} + p \rightarrow n + \nu_{e}\;.     \label{urca2}
 \end{eqnarray}
\end{subequations}
After most of the {\sl ``neutronization''} process is complete, the
standard cooling scenario assumes that neutrino emission proceeds
through the {\sl modified Urca} process:
\begin{equation}
n + n \rightarrow n + p + e^- + \bar\nu_{e} \;.
\label{modified}
\end{equation}
However, given that the modified Urca process requires the presence of
a bystander nucleon to conserve momentum, this process may be millions
of times slower relative to the direct Urca
rate~\cite{Lattimer:2004pg}.  The putative transition into the
significantly slower modified Urca phase is solely based on the
assumption that the proton fraction in the stellar core is below the
$\sim \!15\%$ required to conserve momentum at the Fermi
surface.  However, such an assumption is suspect as the proton
fraction is determined entirely by the density dependence of the
symmetry energy, which remains poorly constrained.  In particular, a
stiff symmetry energy, namely, one that increases rapidly with
density, favors large proton fractions and this may facilitate some
enhanced cooling. Note, however, that unlike other enhanced-cooling
scenarios that may involve the presence of exotic particles in the
core, this enhanced-mechanism is not exotic as it only involves
standard particles and a relatively stiff nuclear symmetry energy.
Also note that recent X-ray observations by the Chandra observatory
seem to suggest that some neutron stars may indeed require some form
of enhanced cooling (see~\cite{Page:2004fy} and references therein).

Given that the proton fraction in neutron-star matter is controlled by
the density dependence of the symmetry energy, a strong correlation
between the onset of the direct Urca process and the neutron-skin
thickness of ${}^{208}$Pb has been established~\cite{Horowitz:2002mb}.
Momentum conservation at the Fermi surface demands that the sum of
the Fermi momenta of the protons and electrons must be greater than
or equal than the neutron Fermi momentum. That is,
\begin{equation}
  k_{\rm F}^{\rm n} \leq  k_{\rm F}^{\rm p} + k_{\rm F}^{\rm e} \;.
\label{UrcaThreshold}
\end{equation}
The onset for the direct Urca process
($k_{\rm F}^{\rm n}\!\equiv\!k_{\rm F}^{\rm p}\!+\!k_{\rm F}^{\rm e}$)
together with the condition of charge neutrality is sufficient
to determine the threshold proton fraction:
\begin{equation}
  Y_{\rm p}^{\rm Urca} =
  \left[1+(1+x_{\rm e}^{1/3})^{3}\right]^{-1}
  \mathop{\longrightarrow}_{x_{\rm e}=1} \,\frac{1}{9} \approx 0.11 \;.
\label{YpThreshold}
\end{equation}
In the above expression $x_{\rm e}$ is the electron-to-proton fraction
and the arrow indicates the model-independent limit of $1/9$ for the
case of a vanishing muon fraction ($x_{\mu}\!=\!1\!-\!x_{\rm
e}\!=\!0$).  For the realistic case of a non-zero muon fraction, the
threshold proton fraction is model dependent and increases slightly to
about $Y_{\rm p}^{\rm Urca}\lesssim 0.15$. In particular, FSUGold
predicts a threshold proton fraction of $Y_{\rm p}^{\rm
Urca}\!=\!0.137$. That is, the FSUGold model predicts that if the
proton fraction within the neutron star exceeds this threshold value,
then enhanced cooling via the direct Urca process is possible. Note
that this threshold proton fraction is reached at a density of about
three times saturation density which corresponds to the central
density of a $M_{\rm Urca}\!=\!1.30\,M_{\odot}$ neutron star (see
Table~\ref{Table5}).  This suggests---in the particular case of the
FSUGold model---that any neutron star with a mass below
$1.30\,M_{\odot}$ that displays enhanced cooling is likely to contain
an exotic core.  For comparison, we should mention that for the
significantly stiffer NL3 equation of state, the threshold proton
fraction ($Y_{\rm p}^{\rm Urca}\!=\!0.129$) is reached at the
significantly lower density of $\rho_{\rm Urca}\!=\!0.205~{\rm
fm}^{-3}$.  This would imply that any neutron star with a mass in
excess of $0.84\,M_{\odot}$ will cool rapidly by the direct Urca
process.  Finally, we note that regardless of the model, the
muon-to-proton fraction at the threshold density is fairly
significant; of the order of 30-40\%.

  \begin{table}[h]
  \begin{tabular}{|c|r|c|}
    \hline
     $A$  & $\langle A\rangle\pm\Delta A\hfil$ & $\rho(A,R_{\rm skin})$ \\
    \hline
     $\rho_{\rm Urca}$  & $(0.4668 \pm 0.1324)\,[28.359\%]$ & $-0.9928$ \\
     $M_{\rm Urca}/M_{\odot}$  & $(1.3012 \pm 0.2658)\,[20.427\%]$ & $-0.9927$ \\
     $Y_{\rm p}^{\rm Urca}$ & $(0.1367 \pm 0.0019)\,[1.421\%]$ & $-0.9927$ \\
     $Y_{\rm p}(2\rho_{0})$ & $(0.1064 \pm 0.0138)\,[13.000\%]$ & $+0.9906$ \\
     $Y_{\rm p}(\rho_{0})$ & $(0.0609 \pm 0.0055)\,[9.055\%]$ & $+0.9166$ \\
     $Y_{\rm p}(\rho_{0}/2)$ & $(0.0346 \pm 0.0051)\,[14.651\%]$ & $-0.9063$ \\
    \hline
  \end{tabular}
 \caption{Theoretical errors associated to the predictions of the
                FSUGold model for various neutron-star properties of
                relevance to the direct Urca process. The threshold
                density is given in units of ${\rm fm}^{-3}$.}
 \label{Table5}
 \end{table}

\begin{figure}[ht]
\vspace{-0.05in}
\includegraphics[width=0.35\columnwidth,angle=0]{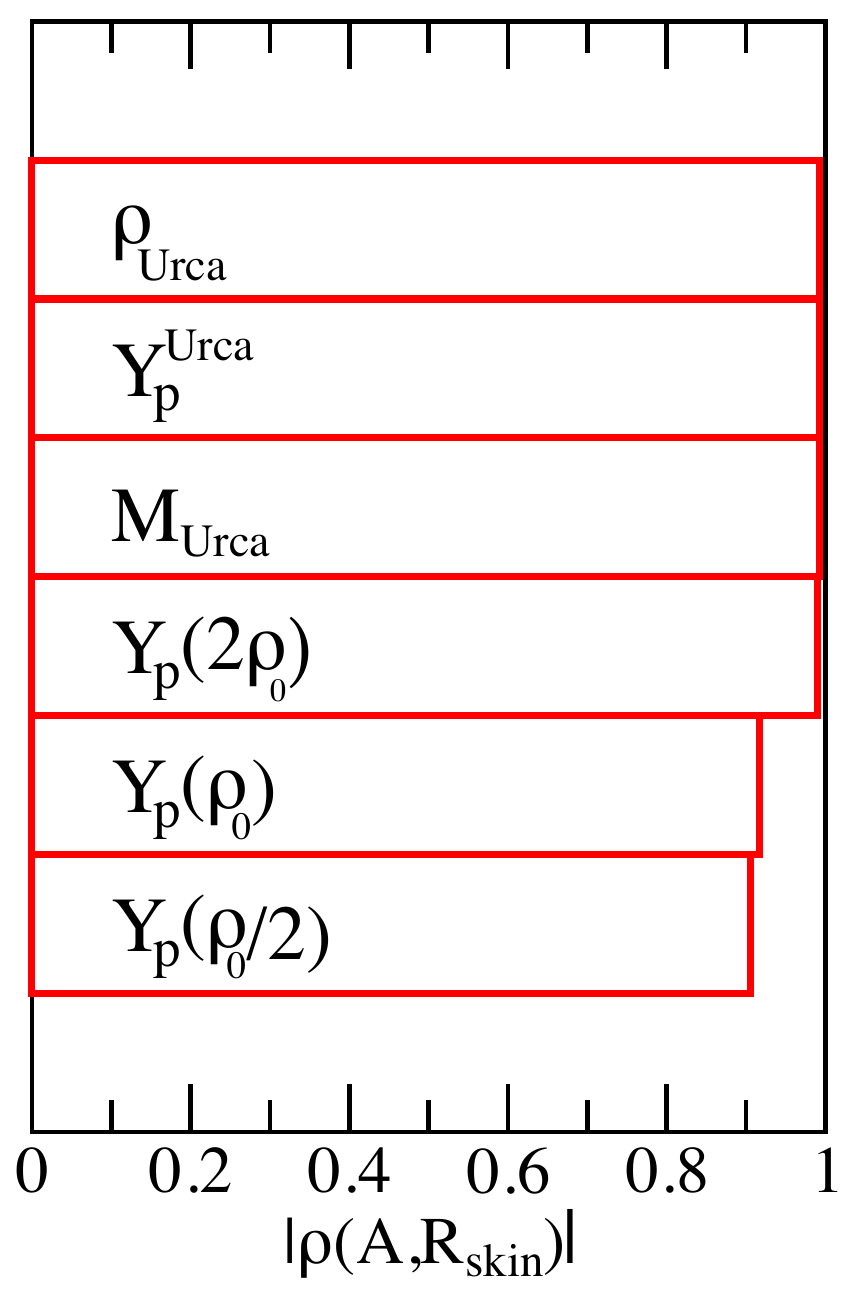}
\caption{(Color online) The correlation coefficient between the
neutron-skin thickness of $^{208}$Pb and various neutron-star
properties of relevance to the direct Urca process.}
\label{Fig5}
\end{figure}

To quantify the strong correlation between various neutron-star
properties at the Urca threshold and the neutron-skin thickness of
${}^{208}$Pb, we have listed in Table\,\ref{Table5} and plotted in
Fig.\,\ref{Fig5} the corresponding theoretical errors and correlation
coefficients extracted from our covariance analysis.  As suggested
above, we find a strong direct correlation between the proton fraction
at high densities and $R_{\rm skin}$. Note, however, that at low
densities the proton fraction and $R_{\rm skin}$ become {\sl
anti}-correlated, as a stiff symmetry energy goes to zero faster than
a soft one. Moreover, as expected, the Urca threshold density, stellar
mass, and proton fraction display a strong (inverse) correlation to
$R_{\rm skin}$.  In particular, we note that the strong
anti-correlation between $Y_{\rm p}^{\rm Urca}$ and $R_{\rm skin}$
emerges entirely because of the presence of muons in the star. Indeed,
without muons $Y_{\rm p}^{\rm Urca}$ is fixed at $1/9$ so all of its
derivatives with respect to the model parameters will vanish and,
thus, so will the correlation coefficient [see
Eq.\,(\ref{Covariance})]. Finally, we have shown that a thin neutron
skin in $^{208}$Pb, {\sl i.e.,} a very soft symmetry energy, requires
a large neutron-star mass for the onset of the direct Urca
process. Thus, we believe that the observation of enhanced cooling in
low-mass neutron stars ($M\!<\!M_{\rm Urca}$) provides one of the most
promising indicators of exotic states of matter residing in the
stellar cores.

\subsection{Core-Crust Transition}
\label{Results:CCT}

Neutron stars have a solid crust above a uniform liquid core.  The
structure and composition of the solid crust remains a source of
significant debate and of considerable
interest~\cite{Baym:1971pw,Ravenhall:1983uh,Hashimoto:1984,
Lorenz:1992zz,Watanabe:2003xu,Watanabe:2004tr,Horowitz:2004yf,
Maruyama:2005vb,Steiner:2007rr,Avancini:2008zz,Newton:2009zz,
Xu:2009vi}. Moreover, the crust is believed to be of critical
importance in a variety of fascinating astrophysical phenomena, such
as pulsar glitches, giant magnetar flares, and the emission of
gravitational waves~\cite{Link:1999ca,Lattimer:2000nx,Steiner:2009yg,
Horowitz:2009ya}.
The phase transition from the solid crust to the uniform liquid core
depends on the properties of neutron-rich matter.  As neutron-rich
matter becomes dilute, the uniform ground state becomes unstable
against small-amplitude density fluctuations~\cite{Horowitz:2000xj,
Carriere:2002bx}. That is, it becomes energetically favorable for the
system to fragment into high-density clusters ({\sl i.e.,} nuclei)
embedded in a dilute neutron-rich vapor. Yet details of the transition
depend sensitively on the proton fraction---and thus on the density
dependence of the symmetry energy---at low densities. Whereas the
proton fraction at high density displays a direct correlation to the
neutron-skin thickness of ${}^{208}$Pb, the proton fraction at low
densities is {\sl anti-correlated} to $R_{\rm skin}$. For example, the
correlation coefficient between $Y_{\rm p}(\rho_{0}/2)$ and $R_{\rm
skin}$ is both large and {\sl negative} (see Table\,\ref{Table5}).
Given that the symmetry energy represents the energy cost in departing
from the isospin symmetric limit, a stiff symmetry energy falls
rapidly to zero at low densities and, hence, tolerates a larger
isospin asymmetry ({\sl i.e.,} smaller $Y_{\rm p}$) than their softer
counterparts. Thus, a lower proton fraction typically implies a low
transition density from the solid crust to the liquid core. This
suggests an inverse correlation: {\sl the thicker the neutron skin
thickness of ${}^{208}$Pb, the lower the core-crust transition
density}~\cite{Horowitz:2000xj}.

  \begin{table}[h]
  \begin{tabular}{|c|r|c|}
    \hline
     $A$  & $\langle A\rangle\pm\Delta A\hfil$ & $\rho(A,R_{\rm skin})$ \\
    \hline
     $P_{\rm t}$  &
     $\quad(0.4020 \pm 0.1071)\,[26.640\%]$ & $+0.9474$ \\
     $Y_{\rm p}^{\rm t}$ &
     $\quad(0.0351 \pm 0.0069)\,[19.711\%]$ & $-0.9260$ \\
     $\mathcal{E}_{\rm t}$ &
     $(71.5337 \pm 5.3747)\,[7.514\%]$ & $-0.9207$ \\
     $\rho_{\rm t}$ &
     $\quad(0.0755 \pm 0.0056)\,[7.369\%]$ & $-0.9203$ \\
    \hline
  \end{tabular}
 \caption{Theoretical errors associated to the predictions of the
                FSUGold model for various neutron-star properties of
                relevance to the transition between the solid crust
                and the uniform liquid core. The transition pressure
                and energy density are given in units of
                ${\rm MeV\,fm}^{-3}$ and the transition baryon density in
                ${\rm fm}^{-3}$.}
 \label{Table6}
 \end{table}

\begin{figure}[ht]
\vspace{-0.05in}
\includegraphics[width=0.35\columnwidth,angle=0]{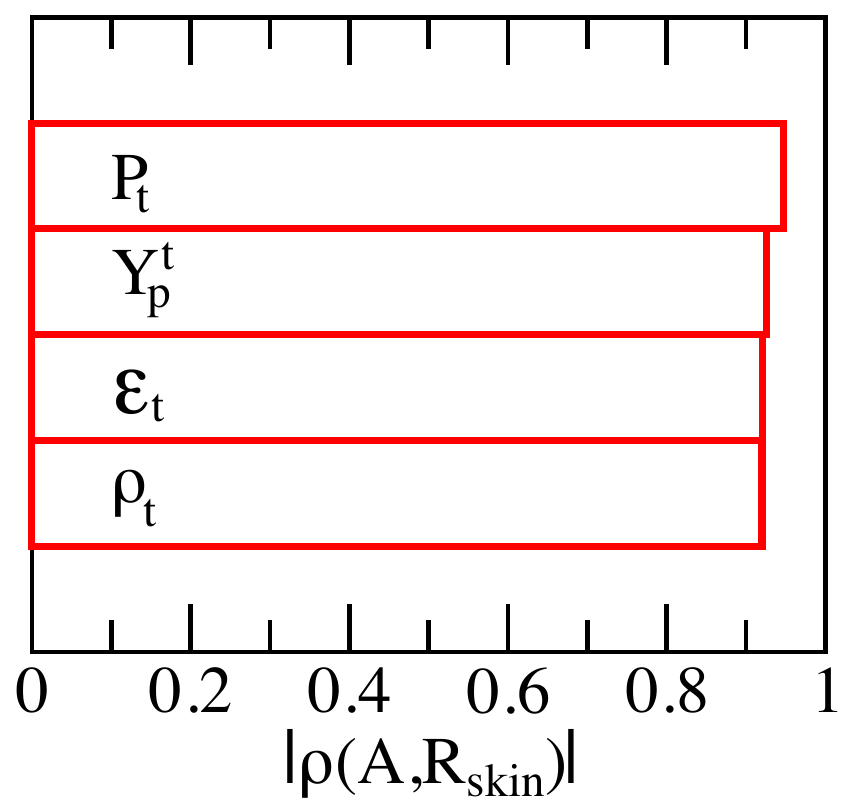}
\caption{(Color online) The correlation coefficient between the
neutron-skin thickness of $^{208}$Pb and various neutron-star
properties of relevance to the transition between the solid crust
and the uniform liquid core.}
\label{Fig6}
\end{figure}

Several approaches---of both a microscopic and thermodynamic
nature---have been used to determine the instability of the uniform
ground-state against cluster formation~\cite{Horowitz:2000xj,
Carriere:2002bx,
Xu:2009vi,Ducoin:2010as,Avancini:2010ch,Ducoin:2011fy}. In the present
study we rely on the RPA stability analysis described in
Ref.~\cite{Carriere:2002bx} to compute the baryon density, proton
fraction, pressure, and energy density at the core-crust
interface. Results from the covariance analysis are listed in
Table\,\ref{Table6} and displayed in Fig.~\ref{Fig6}.
For comparison, the stiffer NL3 equation of state---which has a
significantly smaller symmetry energy at sub-saturation densities than
FSUGold---predicts the following significantly lower values:
$\rho_{t}\!=\!0.052\,{\rm fm}^{-3}$, $Y_{p}^{\rm t}\!=\!0.015$,
$P_{t}\!=\!0.212\,{\rm MeV\,fm}^{-3}$, and
$\mathcal{E}_{t}\!=\!48.960\,{\rm MeV\,fm}^{-3}$. As suggested
earlier, models with a stiffer symmetry energy and thus thicker
neutron skins, display a {\sl ``delayed''} transition from the uniform
core to the solid crust. Thus, we find a strong anti-correlation
between $R_{\rm skin}$ and the various stellar properties at the
core-crust interface~\cite{Horowitz:2000xj}. Yet, the transition
pressure $P_{\rm t}$ behaves in an interesting an unique way. First,
in contrast to other observables, $P_{\rm t}$ appears to be directly
correlated to $R_{\rm skin}$ with a large and positive correlation
coefficient of about $0.95$. Second, the mere existence of a strong
correlation appears to be at odds with earlier studies that suggest a
weak correlation between $P_{\rm t}$ and $R_{\rm
skin}$~\cite{Ducoin:2010as,Ducoin:2011fy,Fattoyev:2010tb}.  We
attribute the present intriguing result to the fact that within the
realm of a covariance analysis, the correlation coefficient between
two observables is obtained by generating model parameters that are
distributed according to the quality measure [see
Eq.\,(\ref{Taylor2})]. That is, models that differ significantly from
the optimal (FSUGold) model carry little weight. In contrast, the weak
correlation between $P_{\rm t}$ and $R_{\rm skin}$ suggested in
Ref.~\cite{Fattoyev:2010tb} was obtained through a systematic (perhaps
even {\sl ad-hoc}) variation around the optimal model.  Although the
covariance analysis implemented here provides the proper statistical
measure of correlation between two observables~\cite{Brandt:1999}, a
covariance analysis can not assess {\sl systematic} errors associated
with the limitations of a given model. Thus, whereas the FSUGold
model---with its soft symmetry energy---predicts a strong positive
correlation between $P_{\rm t}$ and $R_{\rm skin}$, a model with a
stiffer EOS may predict exactly the opposite (see Fig.\,6 of
Ref.~\cite{Fattoyev:2010tb}).

\subsection{Stellar Moment of Inertia}
\label{Results:MOI}

Although the general-relativistic expression for the moment of inertia
of a neutron star is fairly complex---even in the so-called {\sl
slow-rotation approximation}~\cite{Hartle:1967he,Hartle:1968si}---on
simple dimensional grounds it must scale as the product of the
stellar mass times the {\sl square} of its radius. Thus, the
possibility of measuring the moment of inertia to even a 10\% accuracy
may provide stringent constraints on the nuclear equations of
state~\cite{Morrison:2004df,Lattimer:2004nj,Bejger:2005jy,
Lavagetto:2006ew}.  The prospects for such a measurement improved
significantly with the discovery of the binary pulsar PSR
J0737-3039~\cite{Burgay:2003jj,Lyne:2004cj}.  The first ever
discovered double pulsar, PSR J0737-3039 exhibits characteristics that
have enabled the accurate determination of several pulsar properties
(such as the orbital period of the binary, both pulsar masses, and
both spin periods) and have resulted in some of the most precise tests
of Einstein's general theory of relativity.

Motivated by these facts, we have recently studied the sensitivity of
the stellar moment of inertia to the equation of
state~\cite{Fattoyev:2010tb}.  However, we only found a mild
sensitivity of the {\sl total} moment of inertia to the underlying
EOS. Thus, we redirected the main focus of such contribution to the
{\sl crustal} component of the moment of inertia. Several reasons
prompted this choice.  First, constraints on the EOS may be imposed
from an analysis of pulsar glitches in the Vela pulsar that place at
least a 1.4\% of the total moment of inertia in the solid
crust~\cite{Link:1999ca}.  Second, the crust is thin and the density
within it is low, so fairly accurate analytic expressions for the
crustal moment of inertia have been developed in terms of two stellar
properties that are highly sensitive to the EOS: (a) the radius of the
uniform core and (b) the transition pressure at the core-crust
interface~\cite{Link:1999ca,Lattimer:2000nx,Fattoyev:2010tb}.
Finally, given the strong correlation just found between the
core-crust transition pressure $P_{t}$ and $R_{\rm skin}$, one expects
the emergence of a similar strong correlation between the crustal
moment of inertia and $R_{\rm skin}$. Note, however, that our
expectation of a strong correlation is solely based on the results
of the covariance analysis. If instead one relies on a
systematic variations around an optimal model, no strong correlation
was found~\cite{Fattoyev:2010tb}.

  \begin{table}[h]
  \begin{tabular}{|c|r|c|}
    \hline
     $A$  & $\langle A\rangle\pm\Delta A\hfil$ & $\rho(A,R_{\rm skin})$ \\
    \hline
     $I_{0.8}^{\rm cr}$  & $(8.7777 \pm 2.5612)\,[29.178\%]$ & $0.9781$ \\
     $I_{1.4}^{\rm cr}$  & $(5.8988 \pm 1.4055)\,[23.827\%]$ & $0.9619$ \\
     $I_{0.8}$  & $(7.4067 \pm 0.3204)\,[4.326\%]$ & $0.9299$ \\
     $I_{1.4}$  & $(14.7660 \pm 0.3437)\,[2.327\%]$ & $0.5192$ \\
    \hline
  \end{tabular}
 \caption{Theoretical errors associated to the predictions of the
                FSUGold model for the crustal (in units of
                $10^{43}\,{\rm g\,cm}^{2}$) and total (in units of
                $10^{44}\,{\rm g\,cm}^{2}$) moment of inertia of a
                $0.8$ and $1.4$ solar-mass neutron star.}
 \label{Table7}
 \end{table}

\begin{figure}[ht]
\vspace{-0.05in}
\includegraphics[width=0.35\columnwidth,angle=0]{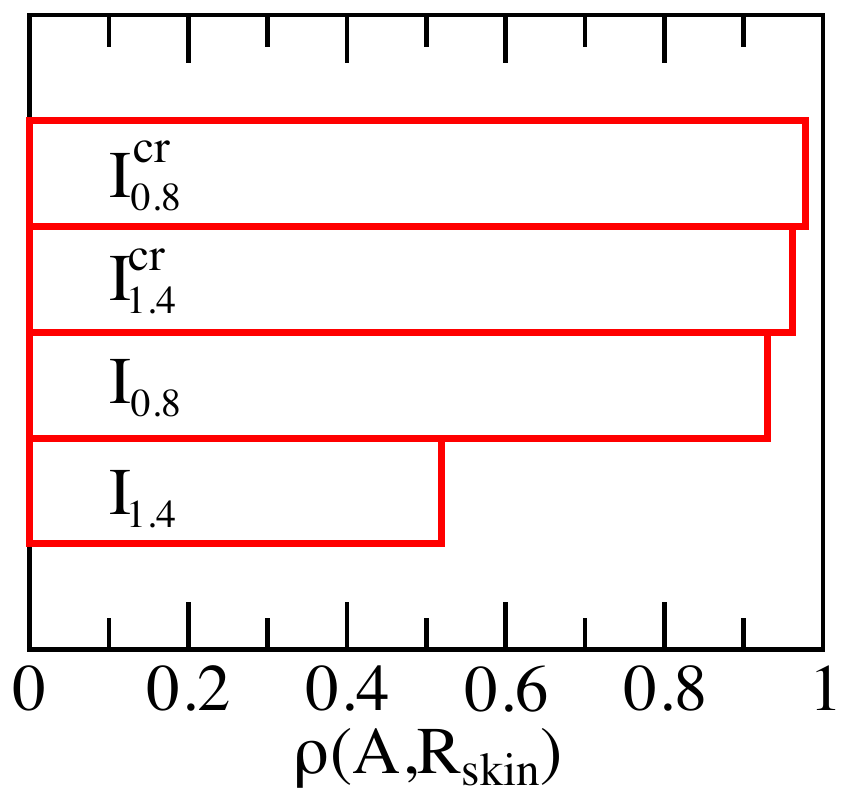}
\caption{(Color online) The correlation coefficient between the
neutron-skin thickness of $^{208}$Pb and various neutron-star
properties of relevance to the crustal and total moment of inertia.}
\label{Fig7}
\end{figure}

In Table\,\ref{Table7} we list the crustal and total moment of inertia
for neutron stars of $0.8$ and $1.4$ solar masses; the correlation
coefficients are also displayed in graphical form in
Fig.\,\ref{Fig7}.  As expected, the crustal moment of
inertia---being sensitive to the core-crust transition pressure
$P_{t}$---displays a strong correlation to $R_{\rm skin}$. Note that
the large error attached to this observable reflects the large
theoretical uncertainty associated with $R_{\rm skin}$---and
ultimately with $L$.  Given that the central density for a low-mass
neutron star is only slightly larger than saturation density, the
correlation between the total moment of inertia of a $0.8\,M_{\odot}$
neutron star and $R_{\rm skin}$ remains strong. However, we
find a significant deterioration in the correlation between $R_{\rm skin}$
and the moment of inertia of a $1.4$ solar-mass neutron star.

\begin{figure}[ht]
\vspace{-0.05in}
\includegraphics[width=1.0\columnwidth,angle=0]{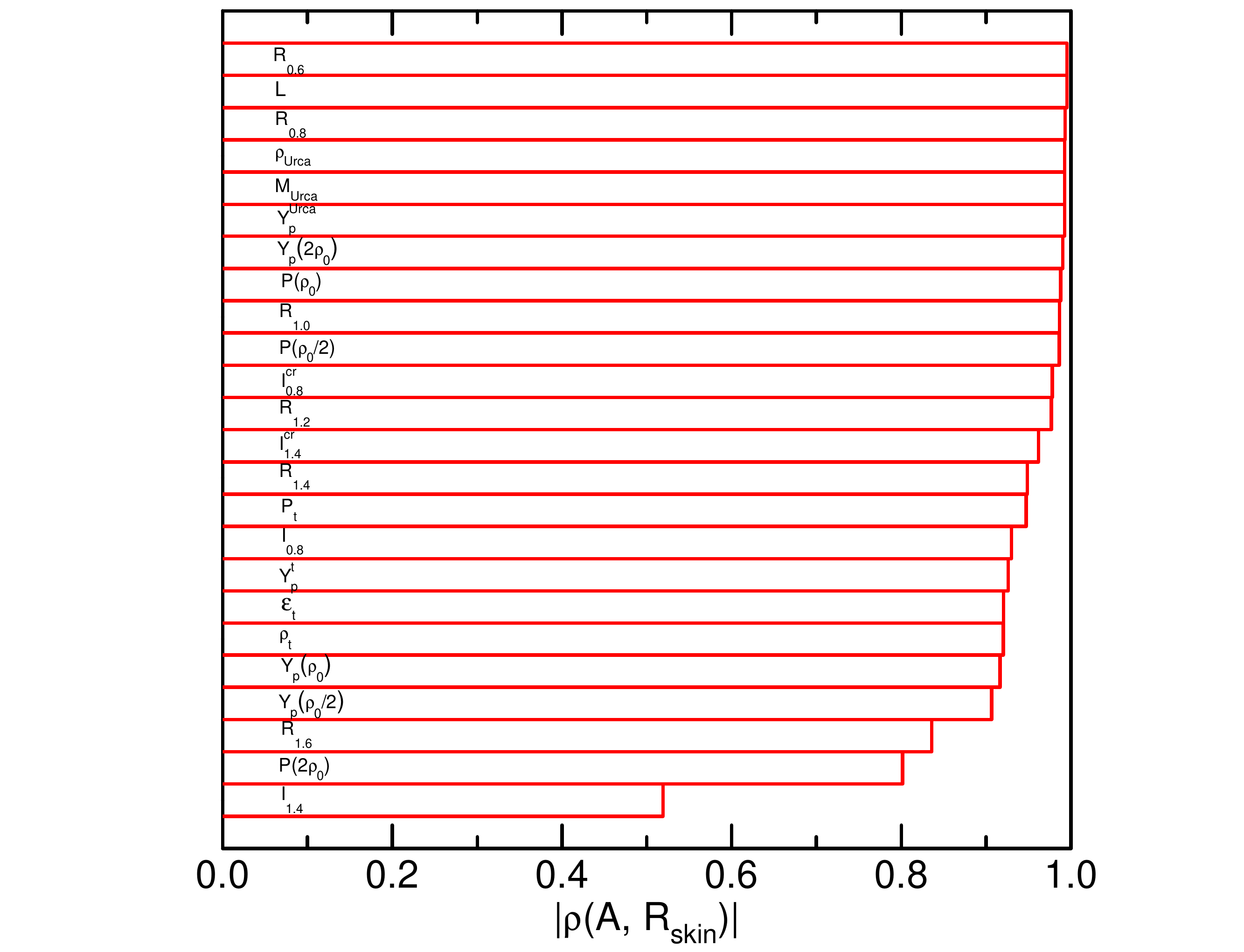}
\caption{(Color online) The correlation coefficient between the
neutron-skin thickness of $^{208}$Pb and the large number of
neutron-star properties discussed in the text.}
\label{Fig8}
\end{figure}

\begin{figure}[ht]
\vspace{-0.05in}
\includegraphics[width=0.6\columnwidth,angle=0]{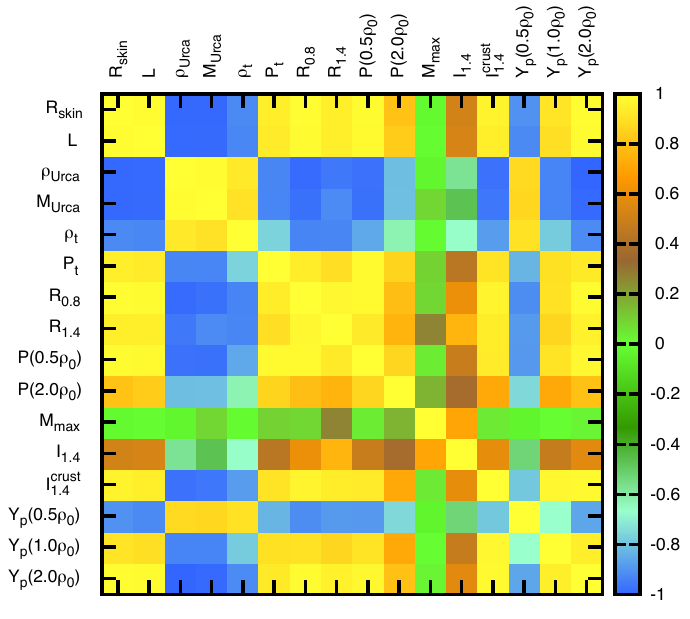}
\caption{(Color online) Color-coded plot of the 120 independent
correlation coefficients between 16 physical observables of
relevance to the structure and dynamics of neutron stars.}
\label{Fig9}
\end{figure}

We close this section by collecting in Fig.\,\ref{Fig8} all
correlation coefficients displayed earlier between $R_{\rm skin}$ and
the large number of neutron-star observables. Moreover, we display in
a color-coded format in Fig.\,\ref{Fig9} correlation coefficients for
16 observables (i.e., 120 independent pairs) of relevance to the
structure, dynamics, and composition of neutron stars. As shown
earlier, most of these stellar observables display a strong
correlation---or anti-correlation---to the neutron skin thickness of
${}^{208}$Pb (first column/row in Fig.\,\ref{Fig9}).  As mentioned
above, a notable exception is the total moment of inertia of a
canonical $1.4\,M_{\odot}$ neutron star. Of course, not every
neutron-star observable is sensitive to the density dependence of the
symmetry energy. The maximum neutron-star mass $M_{\rm max}$, with a
correlation coefficient of only $\rho(M_{\rm max},R_{\rm
skin})\!=\!0.0163$, provides a particularly clear example. Note,
however, that $M_{\rm max}$ yields one the best constraints---if not
the best---on the high density component of the equation of state.

\section{Conclusions}
\label{Conclusions}

The successfully commissioned Lead Radius Experiment (PREX) at the
Jefferson Laboratory has provided the first largely model-independent
determination of neutron-skin thickness of
${}^{208}$Pb~\cite{Abrahamyan:2012gp}: $R_{\rm
skin}\!\equiv\!R_{n}\!-\!R_{p}\!=\!0.33^{+0.16}_{-0.18}~{\rm fm}$.
Given that the determination of the neutron radius of a heavy nucleus
is known to have strong implications on the structure, dynamics, and
composition of neutron
stars~\cite{Horowitz:2000xj,Horowitz:2001ya,Horowitz:2002mb,
Carriere:2002bx,Steiner:2004fi,Li:2005sr,Fattoyev:2010tb}, a detailed
covariance analysis was implemented to quantify the correlations
between $R_{\rm skin}$ and a variety of neutron-star observables.
Moreover, meaningful theoretical uncertainties were provided for all
predicted observables.  We note that although the significant impact
of a measurement of $R_{\rm skin}$ on neutron-star properties has been
known for almost a decade, our work quantifies for the first time
these correlations on the basis of a detailed covariance analysis.  We
stress that the covariance analysis employed here represents the least
biased and most comprehensive tool to uncover correlations between
physical observables~\cite{Reinhard:2010wz,Fattoyev:2011ns}.

In the present study we have computed correlation coefficients between
$R_{\rm skin}$ and the following quantities of direct relevance to the
physics of neutron stars: (a) the equation of state of pure neutron
matter, (b) stellar radii, (c) the onset of the direct Urca proces,
(d) the core-crust phase transition, and (e) the total and crustal
moment of inertia.  In addition to their intrinsic importance, these
observables are interesting as they are sensitive to the density
dependence of the symmetry energy over a wide range of densities. For
example, whereas the onset of the direct Urca process is sensitive to
the symmetry energy at high densities, the core-crust transition
probes the symmetry energy at densities of about a third to a half of
nuclear-matter saturation density.

Although most of the correlations follow the expected trend, two
observables---both strongly correlated to $R_{\rm skin}$---deserve a
special mention. These are the threshold proton fraction for the
direct Urca process $Y_{\rm p}^{\rm Urca}$ and the core-crust
transition pressure $P_{\rm t}$. First, the strong anti-correlation
(of about $-0.99$) between $Y_{\rm p}^{\rm Urca}$ and $R_{\rm skin}$
came as a surprise because the sensitivity of $Y_{\rm p}^{\rm Urca}$
to the density dependence of the symmetry energy is due entirely to
the presence of muons in the stellar core. Indeed, in the absence of
muons $Y_{\rm p}^{\rm Urca}$ is fixed at 1/9 and the correlation
coefficient vanishes. Second, the strong direct correlation between
$P_{\rm t}$ and $R_{\rm skin}$ (of about $+0.95$) is surprising
because no such correlation was observed in previous
studies~\cite{Ducoin:2010as,Ducoin:2011fy,Fattoyev:2010tb}.  Note,
however, that those earlier studies do not rely on a covariance
analysis, but rather explore possible correlations by using either a
large number of nuclear models~\cite{Ducoin:2010as,Ducoin:2011fy}
or models that are systematically varied around an optimal
one~\cite{Fattoyev:2010tb}.  Although the covariance analysis
implemented here provides the proper statistical measure of
correlation between two observables~\cite{Brandt:1999}, a covariance
analysis can not assess {\sl systematic} errors associated with the
limitations of a given model.  Thus, the implementation of a
covariance analysis using other nuclear functionals is both
highly desirable and strongly encouraged.

In connection to the follow-up PREX measurement that aims to achieve
the original 1\% precision goal, we regard it of a great value, as
few accurately calibrated models---if any---predict such a large
(central) value for $R_{\rm skin}$. Moreover, such a large neutron
skin---which would imply a fairly stiff symmetry energy---will have
widespread repercussions in the physics of neutron stars.  It is
worth mentioning, however, that a 10\% uncertainty in the
determination of the slope of the symmetry energy appears to require
a more stringent measurement (at the 0.5\% level) of the neutron
radius of ${}^{208}$Pb~\cite{Fattoyev:2010tb,RocaMaza:2011pm}.
Ultimately, a determination of the density dependence of the
symmetry energy is likely to require a multi-pronged
approach~\cite{Piekarewicz:2007dx}. First, from the theoretical
perspective, we expect that powerful arguments based on the
universality of dilute Fermi gases in the unitary regime will
continue to shed valuable insights into the behavior of pure neutron
matter. Second, in terms of laboratory observables, the dipole
polarizability of ${}^{208}$Pb---measured recently with
unprecedented accuracy~\cite{Tamii:2011pv}---provides a unique
constraint on the density dependence of the symmetry energy and an
excellent complement to the neutron-skin thickness of
${}^{208}$Pb~\cite{Piekarewicz:2012pp}. Finally, enormous advances
in both land- and spaced-based observatories have started to impose
significant constraints on the equation of state.  In this work we
have demonstrated the existence of many neutron-star observables
that---by virtue of their strong correlation---may be used as a
proxy for $R_{\rm skin}$. A particularly interesting alternative is
the radius of a low-mass neutron star, as its central density is
close to nuclear-matter saturation density. Indeed, the radius of
$0.8\,M_{\odot}$ neutron star, $R_{0.8}$, displays a strong
correlation (of $\sim\!0.99$) to $R_{\rm skin}$. Thus, a $5\%$
measurement of $R_{0.8}$ would translate into a $\sim\!0.4\%$
constraint on the neutron radius of ${}^{208}$Pb.  Unfortunately,
the measurement of stellar radii remains a serious challenge that is
further complicated by the scarcity of low-mass neutron stars. Given
that most well measured neutron-star masses lie within a narrow
range centered around $1.4\,M_{\odot}$, low-mass neutron stars may
be difficult to form and therefore may not even exist. In such a
case, a highly accurate measurement of $R_{\rm skin}$ remains the
sole alternative.

\begin{acknowledgments}
This work was supported in part by the United States Department of
Energy under grant DE-FG05-92ER40750 (FSU), by the National
Aeronautics and Space Administration under grant NNX11AC41G issued
through the Science Mission Directorate, and by the NSF grant
PHY-1068022 (TAMUC). We would like to dedicate this work to the
memory of our dear friend and colleague Professor Brian D. Serot.
\end{acknowledgments}

\bibliography{../../ReferencesJP}

\end{document}